\documentclass[12pt]{article}

\usepackage{epsfig}
\usepackage{amssymb}
\usepackage{amsmath}
\usepackage{amsfonts}
       \usepackage{amsthm}

\theoremstyle{theorem}
\newtheorem{lem}{Lemma}[section]

\newtheorem{prop}[lem]{Proposition}

\theoremstyle{definition}

\def\bp{\begin{proof}}
\def\ep{\end{proof}}

\def\be{\begin{equation}}
\def\ee{\end{equation}}
\def\ba{\begin{array}{c}}
\def\baa{\begin{array}{ll}}
\def\ea{\end{array}}

\newcommand{\kt}{\rangle}
\newcommand{\br}{\langle}

\begin{document}

\titlepage

  \begin{center}{\Large \bf

 PT-symmetric model with an interplay
 between kinematical and dynamical non-localities

 }\end{center}

\vspace{5mm}

  \begin{center}

{\bf Miloslav Znojil}

 \vspace{3mm}

Nuclear Physics Institute ASCR, 250 68 \v{R}e\v{z}, Czech
Republic\footnote{ e-mail: znojil@ujf.cas.cz}\\

 \vspace{3mm}

\end{center}

\vspace{5mm}


\section*{Abstract}

A new family of non-Hermitian PT-symmetric quantum models is
proposed in which the Hamiltonians $H=T+V$ are finite-dimensional
and in which the dynamical-input potential $V$ is multi-parametric
and non-local. The choice is supported by the exact solvability of
Schr\"{o}dinger equation and by the well known fact that in
PT-symmetric models a non-locality is already present due to the
generic kinematical non-diagonality of the Hermitizing metrics
$\Theta$. For a subfamily of our $H$s, also {\em all\,} of the
eligible metrics $\Theta$ appear obtainable in closed form.

 \vspace{9mm}%

\noindent PACS

03.65.Ca Formalism

03.65.Fd Algebraic methods

03.65.Aa Quantum systems with finite Hilbert space

\vspace{9mm}

 \noindent MSC 2000:

81Q05, 81Q12,

\vspace{9mm}


\vspace{9mm} KEYWORDS:

non-Hermitian long-range interactions;

discrete quantum Hamiltonians;

closed-form constructions of bound states;

closed-form constructions of
physical inner products;

\newpage

\section{Introduction and summary}

In 1998, Bender and Boettcher \cite{BB} turned attention to ${\cal
PT}-$symmetric (i.e., parity times time-reversal invariant) ordinary
differential operators of the form
 \be
 H=-\frac{d^2}{dx^2} + V(x) \neq H^\dagger
 \label{tradi}
 \ee
and conjectured that in spite of manifest non-Hermiticity, these
operators could still play the role of bound state Hamiltonians for
certain unconventional quantum systems. Almost ten years later Hugh
Jones \cite{Jones} recalled this conjecture (which had been
developed, in between, into a consistent branch of quantum theory
\cite{Carl,ali}) and tried to extend its applicability to
scattering. His results forced him to conclude that one is only
allowed ``to treat the non-Hermitian scattering potential as an
effective one, and work in the standard framework of quantum
mechanics, accepting that this effective potential may well involve
the loss of unitarity'' \cite{Jonesdva}. Fortunately, almost
immediately the threatening crisis has been averted by the
observation \cite{scatt} that the fundamental-theory status of the
whole ${\cal PT}-$symmetric quantum theory (PTSQT) may be
reestablished (i.e., the necessary unitarity of the scattering
process may be reinstalled) via a replacement of the traditional
local-interaction potentials $V(x)$ by their smeared, {\it ad hoc}
non-local forms such that in the coordinate basis $\{\,|x\kt\,\}$
(or rather, for the sake of simplicity, in the discretized
grid-point basis $\{\,|x_j\kt\,\}$) one has $\br x|V|x'\kt \neq 0$,
at some $x \neq x'$ at least.

In our present paper we shall return to the related conceptual
questions and reanalyze the role of non-local interactions in the
bound-state PTSQT context. The key purpose of our study is to
demonstrate, via a family of examples, that one must be very careful
with the use of the traditional concept of locality, for reasons
which were thoroughly explained by Ali Mostafazadeh (cf., e.g., his
papers \cite{ali,alocality}) and which are also summarized briefly
in Appendix A below. Besides this global reminder our specific
toy-model analysis is intended to support the use of dynamically
non-local ${\cal PT}-$symmetric quantum models. Via the description
of our schematic examples we shall show, in particular, that the
introduction of the dynamical non-locality via interaction $V$ need
not spoil the solvability. We shall see that in such a case the
model may prove exactly solvable even in a stronger, PTSQT-related
sense meaning that also the obligatory construction of the physical
inner product (or even of {\it all of the eligible} physical inner
products) may remain feasible by non-numerical means.

The presentation of our results will start in section \ref{sI} where
we shall introduce the terminology and a concrete family of simple
toy models. We shall characterize there the non-locality of a
quantum system as split into its dynamical and kinematical
components. For a large subfamily of our models, moreover, the
kinematical non-locality will be found obtainable, by non-numerical
recurrent means, in explicit form. In section \ref{dvatraja}, such a
constructive kinematics-related result will be then followed by its
dynamics-related parallel in which the wave functions will be found
forming two families, with both of which being obtainable in closed
form. The subsequent section \ref{hya} will offer a deeper,
numerically supported insight into certain characteristic features
of the parameter-dependence and, in particular, of the occurrence of
the domains of reality or, alternatively, of the Kato's \cite{Kato}
exceptional points of complexification of the bound state energies.
Besides the direct localization of the latter points we shall also
emphasize that and how their alternative, indirect localization
could be based on the constructive analysis of the metrics in which
these boundaries of stability of the quantum system acquire the form
of the points of the loss of positivity of the sophisticated
\cite{SIGMA} physical Hilbert-space metric $\Theta \neq I$. Finally,
a few historical and contextual comments will be added in the last
section Nr.~\ref{sV}.

\section{New ${\cal PT}-$symmetric toy-model Hamiltonians \label{sI} }


In Ref.~\cite{scatt}, during the PTSQT-applicability restoration the
decisive progress has been achieved due to a purely technical
simplification based on a non-perturbative replacement of the
continuous axis of $x \in \mathbb{R}$ by its discretized equidistant
version with $x=x_k \sim k$ and $k \in \mathbb{Z}$. This
simplification (which will also be used in what follows and which
may be removed, in principle at least, in the zero-distance grid
point limit) was accompanied by the more or less standard
replacement of the one-dimensional kinetic-energy operator
$T=-d^2/dx^2$ by its difference-operator analogue.

We did -- and also shall -- use the doubly-infinite
tridiagonal-matrix version of the discretized kinetic energy $T$
with elements $T_{kk}=2$ (or, in a shifted-energy regime,
$T_{kk}=0$) and $T_{kk+1}=T_{k+1k}=-1$ in suitable units. Another
assumption of our constructive considerations in \cite{scatt} was
the restriction of attention to the most elementary version of the
non-locality of the interaction $V$. Indeed, in the grid-point
representation  matrix $\br x_k|V|x_m\kt $ contained, in its
simplest version, just the four non-vanishing matrix elements which
lied at the pairs of subscripts $(k,m)=(k_0-1,k_0), (k_0,k_0-1),
(k_0,k_0+1)$ and $(k_0+1,k_0)$. In what follows we shall feel
inspired by the mathematical as well as physical user-friendly
nature of such a four-parametric toy model.

\subsection{Dynamical non-locality}

In a way explained in review paper \cite{ali} the Bender's
physics-inspired requirement of the ${\cal PT}-$symmetry of a
Hamiltonian $H$ (i.e., formal relation ${\cal PT}H = H{\cal PT}$
\cite{Carl}) may be perceived as equivalent to the standard
mathematical requirement
 \be
 {\cal P}H = H^\dagger{\cal P}\,
 \label{PTrule}
  \ee
of the ${\cal P}-$pseudo-Hermiticity {\em alias} Krein-space
Hermiticity of $H$. In other words, the usual time-reversal operator
${\cal T}$ may be perceived as acting on the matrices exemplified by
the potential $\br x_k|V|x_m\kt$ as an antilinear operator of
transposition plus complex conjugation. Concerning the second,
parity-type involution operator ${\cal P}$ such that ${\cal P}^2=I$,
one may decide to make a choice among several standard indefinite
self-adjoint matrix forms of this operator. In our present paper we
shall choose and work with the most common discrete version of the
operator of parity represented by antidiagonal $N$ by $N$
matrix
 \be
 {\cal P}={\cal P}^{(N)}=
 \left [\begin {array}{rrlcrrr}
  &&&&&&1\\{}
  &&&&&\ \ \dot{ \dot{\dot{}\   }\  }&\\{}
  &&&&\  1 &&
 \\{}
 &&& 1 &&&\\
 {}&&1 \ &&&&
 \\
 & {}\ \ \ \
\dot{ \dot{\dot{}\   } \ }&&&&&\\
 1  &&&&&&
 \end {array}\right ]\,.
 \label{PTdef}
 \ee
It is rather straightforward to verify that the requirement of
${\cal PT}-$symmetry allows us to work with the $N=2M+1-$dimensional
Hamiltonians
  \be
 \tilde{H}^{(PT)}=
 \left [\begin {array}{cccc|c|cccc}
 {}{}&-1&{}&{}&{w_{1}}&{}&{}&{}&{}{}{}\\
 {}{}-1&{}& \ddots &{}&{{w_{2}}}&{}&{}&{}{}&{}{}\\
 {}{}{}&\ddots&{}&-1&{\vdots}&{}&{}&{}{}&{} \\
 {}{}{}&{}&-1&{}&w_{M}&{}&{}&{}{}&{}{}\\
 \hline
 {}v_{M}^*&{\ldots}&{{v_{2}^*}} &v_{1}^*
      &u&{w_{M}^*}&\ldots&{{w_{2}^*}}&{w_{1}^*}\\
 \hline
 {}{}{}&{}&{}&{}&v_{1}&{}&-1&{}&{}{} \\
 {}{}{}&{}&{}&{}&{v_{2}}&-1&{}&\ddots&{}{}{}\\
 {}{}{}&{}  &{}&{}&{\vdots}&{}&\ddots&{}&-1{}{}\\
 {}{}{}&{}&{}&{}&{v_{M}}&{}&{}&-1&{}\end {array}\right ]\,
  \label{reforma}
 \ee
which degenerate immediately to their predecessors of
Ref.~\cite{scatt} in the four-parametric special case. The
general, multiparametric version (\ref{reforma}) seems particulary
suitable for our present purposes because it combines a specific
dynamical long-range-interaction non-locality with a formal
simplicity (reflected by the thee-by-three partitioning) and, at
the same time, flexibility (besides a real parameter $u$, the
models vary with as many as $2M$ complex parameters collected in two
$M-$dimensional vectors).

One of the most useful special cases of model (\ref{reforma}) will
be obtained in a maximally asymmetric case with, say, $v_1=-1$ and
trivial subdiagonal long-range couplings $v_2=v_3=\ldots=v_M=0$.
Moreover, the remaining $M-$plet of parameters $w_j$ will be chosen
real. A nontrivial matrix model is then obtained, with
  \be
 \hat{H}^{(MA)}=
 \left [\begin {array}{cccc|c|cccc}
 {}{}&-1&{}&{}&{w_{1}}&{}&{}&{}&{}{}{}\\
 {}{}-1&{}& \ddots &{}&{{w_{2}}}&{}&{}&{}{}&{}{}\\
 {}{}{}&\ddots&{}&-1&{\vdots}&{}&{}&{}{}&{} \\
 {}{}{}&{}&-1&{}&w_{M}&{}&{}&{}{}&{}{}\\
 \hline
 {}&{}&{{}} &-1
      &u&{w_{M}}&\ldots&{{w_{2}}}&{w_{1}}\\
 \hline
 {}{}{}&{}&{}&{}&-1&{}&-1&{}&{}{} \\
 {}{}{}&{}&{}&{}&{}&-1&{}&\ddots&{}{}{}\\
 {}{}{}&{}  &{}&{}&{}&{}&\ddots&{}&-1{}{}\\
 {}{}{}&{}&{}&{}&{}&{}&{}&-1&{}\end {array}\right ]\,
  \label{halfreforma}
 \ee
offering a simplified picture of the long range interaction which is
characterized by its maximal asymmetry.

\subsection{Kinematical non-localities}

In review paper \cite{ali} the author reminded his readers that the
${\cal PT}-$symmetry {\it alias} Krein-space Hermiticity
(\ref{PTrule}) of quantum Hamiltonians $H$ plays in fact just a
useful but purely auxiliary heuristic role. The correct physical
interpretation may only be deduced from the hidden-Hermiticity
relations
 \be
 H^\dagger \Theta=\Theta\,H\,
 \label{nelamb}
 \ee
in which a suitable operator $\Theta$ of a physical Hilbert space
metric is positive definite (see a compact review of the whole
formalism as well as a more detailed discussion of this point in
Appendix A below).

Naturally, with the trivial choice of $\Theta^{(Dirac)}=I$ one would
immediately return back to the conventional Hermiticity requirement
$H^\dagger=H$ of textbooks. In particular, in a way illustrated by
the purely kinetic choice of $H_0=T=-d^2/dx^2=H_0^\dagger$ one could
then still speak about a ``kinematical non-locality" related, in the
above discrete-coordinate picture, to the ``residual
non-diagonality'' of matrices (\ref{reforma}) or (\ref{halfreforma})
even in the complete absence of interaction couplings $v_j$ and
$w_k$.

In what follows, we shall strictly speak about a ``kinematical
non-locality'' in another sense in which the genuinely
non-dynamical, potential-independent non-locality becomes introduced
via the non-diagonal-matrix nature of the inner-product metric
$\Theta \neq I$ (cf. Appendix A once more). In such a setting, an
important consequence of the choice of special toy model
(\ref{halfreforma}) with a nontrivial dynamical part lies in the
related thorough simplification of Eq.~(\ref{nelamb}). Purely
formally, these relations may be then interpreted as an
underdetermined linear algebraic system of equations
 \be
 {\cal M}_{ij}=\left (H^\dagger \Theta-\Theta\,H
 \right )_{ij}=0\,,
 \ \ \ \ \ \ i,j=1,2,\ldots,N\,
 \label{dieuy}
 \ee
which define all of the eligible metrics which would make a given
Hamiltonian (\ref{halfreforma}) self-adjoint in the related
sophisticated physical Hilbert space ${\cal H}^{(S)}$. Indeed, the
main reason for our restriction of attention to matrix
$\hat{H}^{(MA)}$ which is so extremely non-Hermitian in the friendly
but false Hilbert space ${\cal H}^{(F)}=\mathbb{R}^N$ lies in the
emergent and unexpected feasibility of construction of {\em all} of
the related eligible Hermitizing metrics at {\em arbitrary} matrix
dimensions $N=2M+1$.

\begin{prop}

For the whole $(M+1)-$parametric family of our real and maximally
asymmetric $N=(2M+1)-$dimensional and ${\cal PT}-$symmetric
Hamiltonian matrices (\ref{halfreforma}) the construction of the
necessary Hermitian (i.e., real and symmetric)
candidates for a positive-definite Hermitizing metric
  \be
 \Theta^{(candidate)}=
  \left[ \begin {array}{cccc}
  \Theta_{11}&\Theta_{12}&\ldots&\Theta_{1N}\\
  \noalign{\medskip}
  \Theta_{12}&\Theta_{22}&\ldots&\Theta_{2N}\\
 \noalign{\medskip}\vdots&\vdots&&\vdots
 \\
  \noalign{\medskip}
  \Theta_{1N}&\Theta_{2N}&\ldots&\Theta_{NN}
 \end {array} \right]
 \label{thehami5}
 \ee
may proceed via recurrent solution of linear  algebraic system
(\ref{nelamb}).

\end{prop}

\bp

Once one starts from Eq.~(\ref{dieuy}) and decides to treat the
first row of matrix $\Theta^{(candidate)}$ as the $N-$plet of
independent variable parameters, the closed-form recurrent
construction of the further matrix elements may proceed row-wise,
leading to the exhaustive sequence of definitions of
 \be
 \Theta_{22}\,,\Theta_{23}\,,\ldots\,,\Theta_{2N}\,,
 \Theta_{33}\,,\Theta_{34}\,,\ldots\,,\Theta_{NN}\,
 \label{rec1}
 \ee
which are provided, respectively, by the independent linear
equations
 \be
 {\cal M}_{12}=0\,,
 {\cal M}_{13}=0\,,\ldots\,,{\cal M}_{1N}=0\,,
 {\cal M}_{23}=0\,,{\cal M}_{24}=0\,,\ldots\,,{\cal M}_{N-1N}=0\,
 \label{rec2}
 \ee
ordered and selected out of the redundant system (\ref{dieuy}). This
property is a consequence of the extremely friendly sparsity of the
Hamiltonian. Due to the Hermiticity of matrix ${\cal M}$, the
verification of the recurrent-relation correspondence between the
linear equations (\ref{rec2}) and their solutions
(\ref{rec1}) is provided by the insertion of $H$ and of
the Hermitian-matrix ansatz (\ref{thehami5}) in the system of $N^2$
linear algebraic equations~(\ref{dieuy}).

\ep

Let us add a  remark that the possibility of using the first row of
matrix $\Theta^{(candidate)}$ as free parameters implies that one
may treat all metrics as linear superpositions of their
simpler-matrix components. It makes sense to emphasize here that
these components are in general different from the components
provided by
%
%
%
the well known alternative spectral-like expansion of the metric
\cite{SIGMAdva},
 \be
 \Theta=\Theta( \kappa^2_1, \kappa^2_2,\ldots, \kappa^2_{2M+1})
 =\sum_{j=1}^{2M+1}\,|\Psi_j\kt \kappa^2_j\,\br \Psi_j|\,.
 \label{famu}
 \ee
In the real- and non-degenerate-spectrum case the necessary
$(2M+1)-$plet of input vectors $|\Psi_j\kt \in {\cal H}^{(F)}$ must
be made available here in the form of a complete set of
non-orthogonal, arbitrarily normalized eigenvectors of the Hermitian
conjugate of our non-Hermitian input Hamiltonian,
 \be
 \left ( \tilde{H}^{(PT)} \right
 )^\dagger\,|\Psi_j\kt=\varepsilon_j\,|\Psi_j\kt\,,
 \ \ \ \ \ j=1,2,\ldots,2M+1\,.
 \ee
The techniques of solution of such a conjugate form of
Schr\"{o}dinger equation are to be discussed below.

\section{Wave functions\label{dvatraja}}

\subsection{Right eigenvectors of the Hamiltonian}
%
%

Our tilded toy-model Hamiltonians (\ref{reforma}) are
triply partitioned,
  \be
 \tilde{H}^{(PT)}=
 \left [\begin {array}{c|c|c}
 D&\vec{w}&0\\
 \hline
 \left (\vec{v}\right )^\dagger {\cal P}^{(M)}
      &u&\left (\vec{w}\right )^\dagger{\cal P}^{(M)}\\
      \hline
 0&\vec{v}&D
 \end {array}\right ]\,.
 \label{trireforma}
 \ee
The real and symmetric tridiagonal $M$ by $M$ submatrix $D$ has the
non-degenerate real spectrum \cite{ptsqw} and
obeys the symmetry relation  $D =  {\cal P}^{(M)} D {\cal P}^{(M)}$.
In terms of a unitary $M$ by $M$ matrix $U$ with known,
Chebyshev-polynomial elements \cite{Cheby} this matrix may be diagonalized
non-numerically,
$D =
U^\dagger_D \hat{d} U_D$. The new diagonal
matrix $\hat{d}$ is such that $\hat{d} {\cal P}^{(M)}+{\cal
P}^{(M)}\hat{d} =0$. Hence, we may replace our
Hamiltonian $\tilde{H}^{(PT)}$ by its
simpler, untilded form
$ H^{(PT)} = {\cal U} \tilde{H}^{(PT)} {\cal U}^\dagger\,
$
where
 \be
{\cal U}=\left [\begin {array}{c|c|c}
 U_D {\cal P}^{(M)}&0&0\\
 \hline
 0
      &1&0\\
      \hline
 0&0&U_D
 \end {array}\right ]
\,.
 \label{trireformabe}
 \ee
%
%
%
%
Using abbreviations $\vec{\alpha}=U_D{\cal P}^{(M)}\vec{w}$ and
$\vec{\beta}=U_D\vec{v}\, $, the $(2M+1)-$plet of
bound states of our present quantum model becomes now
defined by the partitioned Schr\"{o}dinger equation
 \be
 H^{(PT)} \,
 \left [\begin {array}{c}
 \vec{x}\\
 z\\
 \vec{y}
 \end {array}\right ]=\varepsilon\,
 \left [\begin {array}{c}
 \vec{x}\\
 z\\
 \vec{y}
 \end {array}\right ]\,,
 \ \ \ \ \ \  H^{(PT)} =
 \left [\begin {array}{c|c|c}
 \hat{d}&\vec{\alpha}&0\\
 \hline
 \left (\vec{\beta}\right )^\dagger
      &u&\left (\vec{\alpha}\right )^\dagger\\
      \hline
 0&\vec{\beta}&\hat{d}
 \end {array}\right ]\,.
 \label{SEright}
 \ee
The process of its solution splits into two parts. Firstly, we
assume that $z=0$. This reduces the problem to the two trivially
solvable diagonal-matrix subproblems $(\hat{d}-\varepsilon\,\hat{I})
\vec{x}=0$ and  $(\hat{d}-\varepsilon\,\hat{I}) \vec{y}=0$. Hence,
it is easy to prove

  \begin{prop}

The first, $z=0$ subset of solutions of Schr\"{o}dinger
Eq.~(\ref{SEright}) is an $M-$plet such that its energies
$\varepsilon_j^{(z=0)}=\hat{d}_j$ are real and independent of
parameters $\vec{\alpha}$ and $\vec{\beta}$. For wave functions we
have $\vec{x}_m=\vec{y}_m=0$ at all $m \neq j$ for all
$j=1,2,\ldots,M$. The remaining middle-line constraint
$\vec{\beta}^*_j\,\vec{x}_j+\vec{\beta}^*_j\,\vec{y}_j=0$ just makes
the first set of wave functions unique after (arbitrary)
normalization.

 \end{prop}

For the ``missing'' second set of  $M+1$ bound-state solutions we
may select $z=1$ which fixes their norm. We shall skip the detailed
discussion of exceptional cases and assume merely, for the sake of
simplicity, that $\varepsilon \neq \hat{d}_m$ at all
$m=1,2,\ldots,M$. From our Schr\"{o}dinger Eq.~(\ref{SEright}) we
may then extract the explicit definition of the wave functions,
 $$
 \vec{x}=-\frac{1}{\hat{d}-\varepsilon\,\hat{I}}\,\vec{\alpha}\,,\ \ \ \
 \vec{y}=-\frac{1}{\hat{d}-\varepsilon\,\hat{I}}\,\vec{\beta}\, .
 $$
Ultimately, the remaining middle-row algebraic-equation remnant
 \be
 \varepsilon_j=u+R(\varepsilon_j)\,,
 \ \ \ \
 R(\varepsilon_j)=
 \sum_{i=1}^{M}\,\left (
 \beta_i^*\frac{1}{\varepsilon_j-\hat{d}_i}\,\alpha_i+
 \alpha_i^*\frac{1}{\varepsilon_j-\hat{d}_i}\,\beta_i
 \right )\,,\ \ \  j = 1, 2, \ldots, M\,
 \label{rooty}
 \ee
of our Schr\"{o}dinger equation determines the spectrum. Although
the explicit solution of the latter equation is, in general, a
purely numerical problem, we immediately see that our auxiliary
function $R(\varepsilon)$ is real and, up to its singularities at
the first subset of energies $\varepsilon=\hat{d}_i$, continuous,
with elementary asymptotics
 $$
 R(\varepsilon) = \frac{G}{\varepsilon}+{\cal O}
 \left ( \frac{1}{\varepsilon^2}\right )\,,
 \ \ \ \
 G=\sum_{i=1}^{M}\,\left (
 \beta_i^*\,\alpha_i+
 \alpha_i^*\,\beta_i
 \right )\,.
 $$
Moreover, up to the possible exceptional degenerate cases, the
singularities are real and isolated first-order poles. This
observation completes the proof of the following result.

   \begin{prop}


The second, $z=1$ set (i.e., the ``missing'' $(M+1)-$plet) of the
$\vec{\alpha}-$ and $\vec{\beta}-$dependent bound-state energy roots
$\varepsilon_m$ is defined by the transcendental secular equation
(\ref{rooty}). Up to the above-mentioned exceptional degenerate
cases these roots are real and ordered,
 \be
 \varepsilon_1<\hat{d}_1<\varepsilon_2<\ldots
 < \hat{d}_M<\varepsilon_{M+1}\,,
 \ee
{\it i.e.}, separated by the poles of $R(\varepsilon)$, {\it i.e.},
by the remaining energy levels belonging to the first, $z=0$ set.

   \end{prop}

\subsection{Left eigenvectors of the Hamiltonian}

A remark is to be added now concerning the parallels between the
Schr\"{o}dinger's bound-state problem (\ref{SEright}) and
its conjugate forms needed in preceding section. Naturally, the
partitioning may be recalled in conjugate case yielding the relations
 \be
 \label{SEleft}
 \left [\begin {array}{c|c|c}
 \hat{d}&\vec{\beta}&0\\
 \hline
 \left (\vec{\alpha}\right )^\dagger
      &u&\left (\vec{\beta}\right )^\dagger\\
      \hline
 0&\vec{\alpha}&\hat{d}
 \end {array}\right ]
 \,
 |\Psi_j\kt
 =\varepsilon_j\,
 |\Psi_j\kt\,,\ \ \ \ \ \
 |\Psi_j\kt=
  \left [\begin {array}{c}
 \vec{X}\\
 Z\\
 \vec{Y}
 \end {array}\right ]\,.
 \ee
The solution remains analogous to the non-conjugate case
so we need not describe it in detail. It is only
worth mentioning that the treatment of the new, upper-case Schr\"{o}dinger
equation (\ref{SEleft}) will be facilitated by our knowledge of the spectrum.
The comparison of the lower- and upper-case
wave functions reveals that the left and right eigenstates of our
Hamiltonians are also closely related by an elementary interchange
of the two vectors of the
complex parameters
$\vec{\alpha} \leftrightarrow \vec{\beta}$.

One could summarize that at any $M$, the closed-form construction of
the necessary physical Hermitizing metric via expansion~(\ref{famu}) is
only marginally more complicated than the explicit specification of
the bound states themselves. At the same time, the mutual
non-orthogonality of vectors $|\Psi_j\kt$ makes the resulting
matrices of the metric cumbersome. Even at the smallest integers $M$,
they can hardly be displayed in print, therefore.


On positive side,
a definite advantage of using formula (\ref{famu}) lies in the
fact that the domain of the acceptable parameters $\kappa_j^2$
keeping the metric positive definite is trivial. In contrast, the
use of the recurrent construction of Proposition 2.1 does not
provide any sufficiently general guidance for the choice and/or
limitations of variability of the free parameters forming the first
row of the real matrix $\Theta^{(candidate)}$. A
case-by-case analysis is usually needed.

\section{Numerical results\label{hya}}

\subsection{The domains of reality of the
energies\label{hyasd}}

%
%
%
%

\begin{figure}[h]                     
\begin{center}                         
\epsfig{file=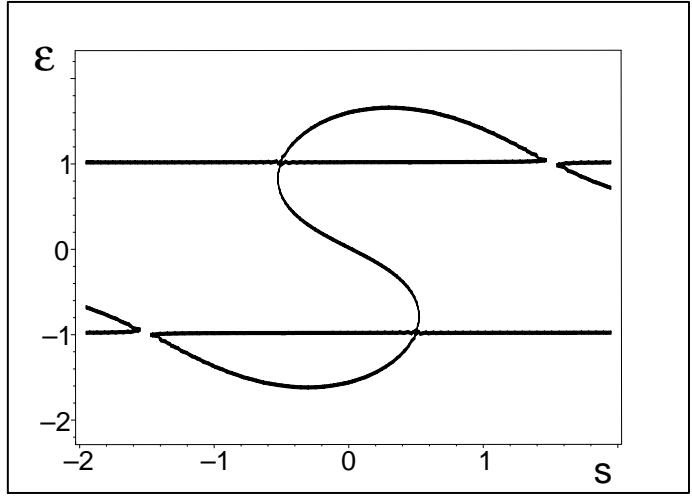,angle=270,width=0.6\textwidth}
\end{center}                         
\vspace{-2mm} \caption{The sample of $s-$dependence of the real
eigenvalues of Hamiltonian $\tilde{H}^{(PT)}(r,s)$ of
Eq.~(\ref{hami5}) at $r=1/2$. \label{firm}}
\end{figure}

Let us start by considering the first nontrivial $M=2$ (i.e.,
five-by-five) matrix (\ref{reforma}) in which we assume, for the
sake of simplicity, that the interaction part is just real and
antisymmetric. This yields the two-parametric toy-model Hamiltonian
 \be
 \tilde{H}^{(PT)}(r,s)=
  \left[ \begin {array}{ccccc}
  0&-1&r&0&0\\\noalign{\medskip}-1&0&-1+s&0
&0\\\noalign{\medskip}-r&-1-s&0&-1+s&r\\\noalign{\medskip}0&0&-1-s&0&-
1\\\noalign{\medskip}0&0&-r&-1&0\end {array} \right] \label{hami5}
 \ee
in which we further set $r=1/2$. Then,  secular equation
 $$
 {{\it \varepsilon}}^{5}-7/2\,{{\it \varepsilon}}^{3}+2\,{{\it
 \varepsilon}}^{3}{s}^{2}-2\,s{{
 \it \varepsilon}}^{2}+5/2\,{\it \varepsilon}-2\,{\it \varepsilon}\,{s}^{2}+2\,s=0
 $$
may be solved exactly. Its solution determines the roots which
remain all real in an interval of $s \in (-0.5242\ldots,
0.5242\ldots)$ while a pair of these roots merges and becomes
complex everywhere out of such an interval (cf. Fig.~\ref{firm}).

Although even a very sketchy inspection of Fig.~\ref{firm}
immediately confirms that the two spectral-reality-boundary points
$s =\pm 0.5242\ldots$ are complexification singularities, i.e.,
exceptional points (EPs) in the sense of definition as given by Kato
\cite{Kato}, one may feel truly puzzled by the very small distance
of these boundary points from the other two points of slightly
smaller absolute value at which the curve happens to cross the
constant-root lines of $ \varepsilon_\pm = \pm 1$.

A deeper analysis of these intersections localizes them at $s=\pm
1/2$. Moreover, their determination in closed form enabled us to
arrive at a truly surprising observation that these points in the
interior of the interval of the spectral reality are no ``normal
crossings with the two straight lines'' but rather that they form
another pair of the other two Kato's EP singularities. In other
words, one reveals that for the same toy model the values of $s=\pm
1/2$ at which the respective pairs of the real eigenvalues collide
are {\em also} the points at which our matrix
$\tilde{H}^{(PT)}(1/2,s)$ ceases to be diagonalizable and, hence,
acceptable as a Hamiltonian operator in quantum mechanics. Thus, one
arrives at the following counterintuitive, interesting and truly
important observation.

  \begin{prop}

For our illustrative $M=2$ Hamiltonian $\tilde{H}^{(PT)}(1/2,s)$ of
Eq.~(\ref{hami5}) the physical domain of its single free parameter
$s$ splits in three open intervals, viz.,  ${\cal
D}_-=(-s_{EP},-1/2)$, ${\cal D}_0=(-1/2,1/2)$ and ${\cal
D}_+=(1/2,s_{EP})$ with an empty overlap. In terms of an
abbreviation $f=\sqrt [3]{5+\sqrt {33}}$ we also obtain the exact
and closed formula
 \be
 s_{EP}={\frac {   {2}\sqrt {6\,{f}^{2}-15\,f-f
\sqrt {33}+21+5\,\sqrt {33}}-2\,\sqrt{2}\,f  }{4\,\sqrt {f({
f}^{2}-2)}}}
 \approx 0.5242106130\ldots\,
 \label{clofo}
 \ee
determining our initial, reality-boundary EPs.

  \end{prop}

 \bp

 \noindent
The derivation of closed formula (\ref{clofo}) is based on the
secular equation re-interpreted as a quadratic equation for two
functions $s=s_\pm(\varepsilon)$.
%
%
For a completion of the proof the EP property of $s=\pm 1/2$ must be
verified via the direct evaluation of the related eigenvectors. At
either of the two doubly degenerate eigenvalues one really obtains
just a single eigenvector (with, e.g., $x_1=x_2=z=0$ and $y_1=y_2=1$
at $s=1/2$ and $\varepsilon=-1$, etc). Thus, the basis in our friendly Hilbert
space ${\cal H}^{(F)}=\mathbb{R}^5$ must be completed by the fifth,
``missing'' associated-vector element (with, e.g., $x_1=0$,
$x_2=-1/2$, $z=-1$, $y_1=0$ and $y_2=1/2$  at $s=1/2$ and $\varepsilon=-1$,
etc).

 \ep


\begin{figure}[h]                     
\begin{center}                         
\epsfig{file=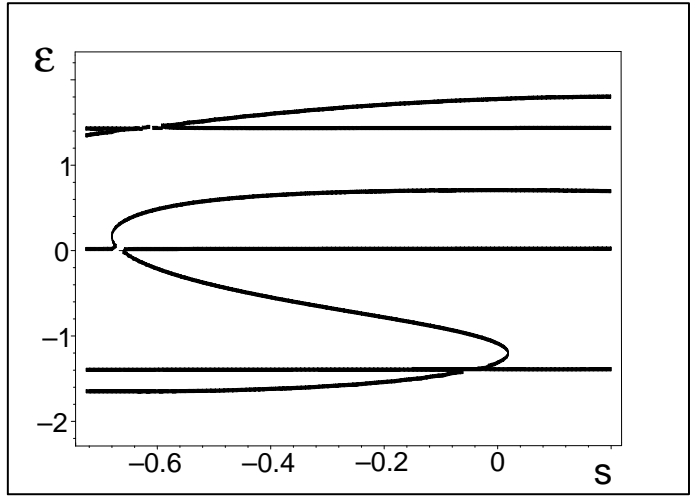,angle=270,width=0.6\textwidth}
\end{center}                         
\vspace{-2mm} \caption{The sample of $s-$dependence of the real
eigenvalues of Hamiltonian $\tilde{H}^{(PT)}(q,r,s)$ of
Eq.~(\ref{hami7}) at $r=1/2$ and $q=1/3$. \label{fidm}}
\end{figure}



The overall pattern does not vary with the dimension $M$ too much.
Thus, in the $M=3$ toy-model Hamiltonian
  \be
 \tilde{H}^{(PT)}(q,r,s)=
 \left[ \begin {array}{ccccccc}
 0&-1&0&q&0&0&0\\
 \noalign{\medskip}-1&0
 &-1&r&0&0&0\\
 \noalign{\medskip}0&-1&0&-1+s&0&0&0\\
 \noalign{\medskip}-q
 &-r&-1-s&0&-1+s&r&q\\
 \noalign{\medskip}0&0&0&-1-s&0&-1&0
 \\
 \noalign{\medskip}0&0&0&-r&-1&0&-1\\
 \noalign{\medskip}0&0&0&-q&0&-1&0
 \end {array} \right]
 \label{hami7}
 \ee
we may set, for illustration purposes, $r=1/2$ and $q=1/3$ and get
the spectrum sampled in Fig.~\ref{fidm}. It is worth noticing that
due to the presence of four EPs, the physical domain ${\cal D}$ (in
which all energies remain real) is now split into four
non-overlapping subdomains. In each of them the set of energies is
formed by an $s-$independent triplet and an $s-$dependent quadruplet
-- in a way predicted in section \ref{dvatraja}.



\begin{figure}[h]                     
\begin{center}                         
\epsfig{file=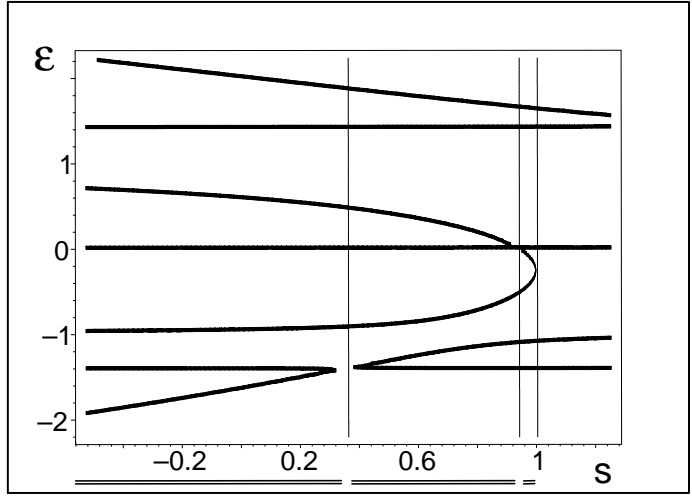,angle=270,width=0.6\textwidth}
\end{center}                         
\vspace{-2mm} \caption{The three ``twice-underlined'' $s-$domains of
the reality of all of the seven eigenvalues of the asymmetric
Hamiltonian of Eq.~(\ref{hami27}) at $r=1/2$ and $q=-1/15$.
\label{fidm715}}
\end{figure}

\begin{figure}[h]                     
\begin{center}                         
\epsfig{file=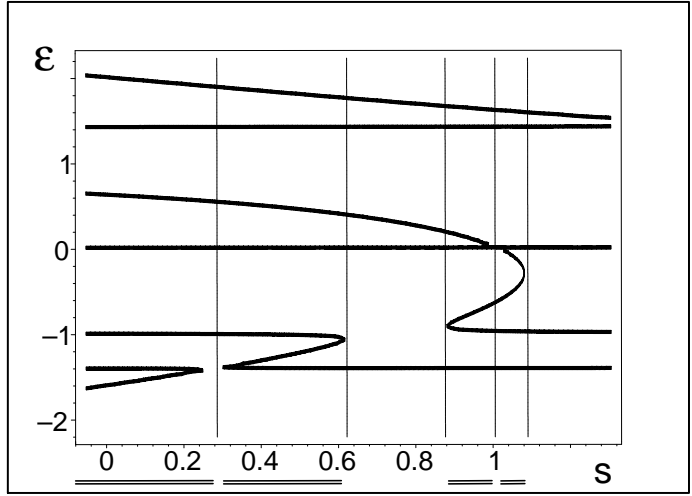,angle=270,width=0.6\textwidth}
\end{center}                         
\vspace{-2mm} \caption{The emergence of a gap between the two pairs
of domains of the full reality of spectrum for
Hamiltonian~(\ref{hami27}) at $r=1/2$ and $q=1/100$.
\label{fidm7100}}
\end{figure}


\subsection{A gap between domains}

{With the asymmetry of interaction made maximal} let us now apply
our numerical sampling also to model (\ref{halfreforma}). As long as
we now only have $\beta=U_D\vec{v} \neq \vec{0}$, it makes sense to
start our discussion from the choice of $M=3$ with $u=0$ yielding
the asymmetric toy-model Hamiltonian
 \be
 \hat{H}^{(PT)}(q,r,s)=
  \left[ \begin {array}{ccccccc} {}&-1&{}&q&{}&{}&{}\\\noalign{\medskip}-1&{}
&-1&r&{}&{}&{}\\\noalign{\medskip}{}&-1&{}&-1+s&{}&{}&{}\\\noalign{\medskip}{}&{}
&-1&{}&-1+s&r&q\\\noalign{\medskip}{}&{}&{}&-1&{}&-1&{}\\\noalign{\medskip}{}
&{}&{}&{}&-1&{}&-1\\\noalign{\medskip}{}&{}&{}&{}&{}&-1&{}\end
{array} \right]\,.
 \label{hami27}
 \ee
First of all, one has to analyze the structure of the physical
parametric domain ${\cal D}$ and, in particular, of its EP
boundaries $\partial {\cal D}$. For our present purposes, let us
just mention that whenever we follow our preceding strategy and
restrict the role of a variable parameter, say, to $s$, the
$s-$dependence of the real eigenvalues $\varepsilon_j$ remains
qualitatively the same as above.


\begin{figure}[h]                     
\begin{center}                         
\epsfig{file=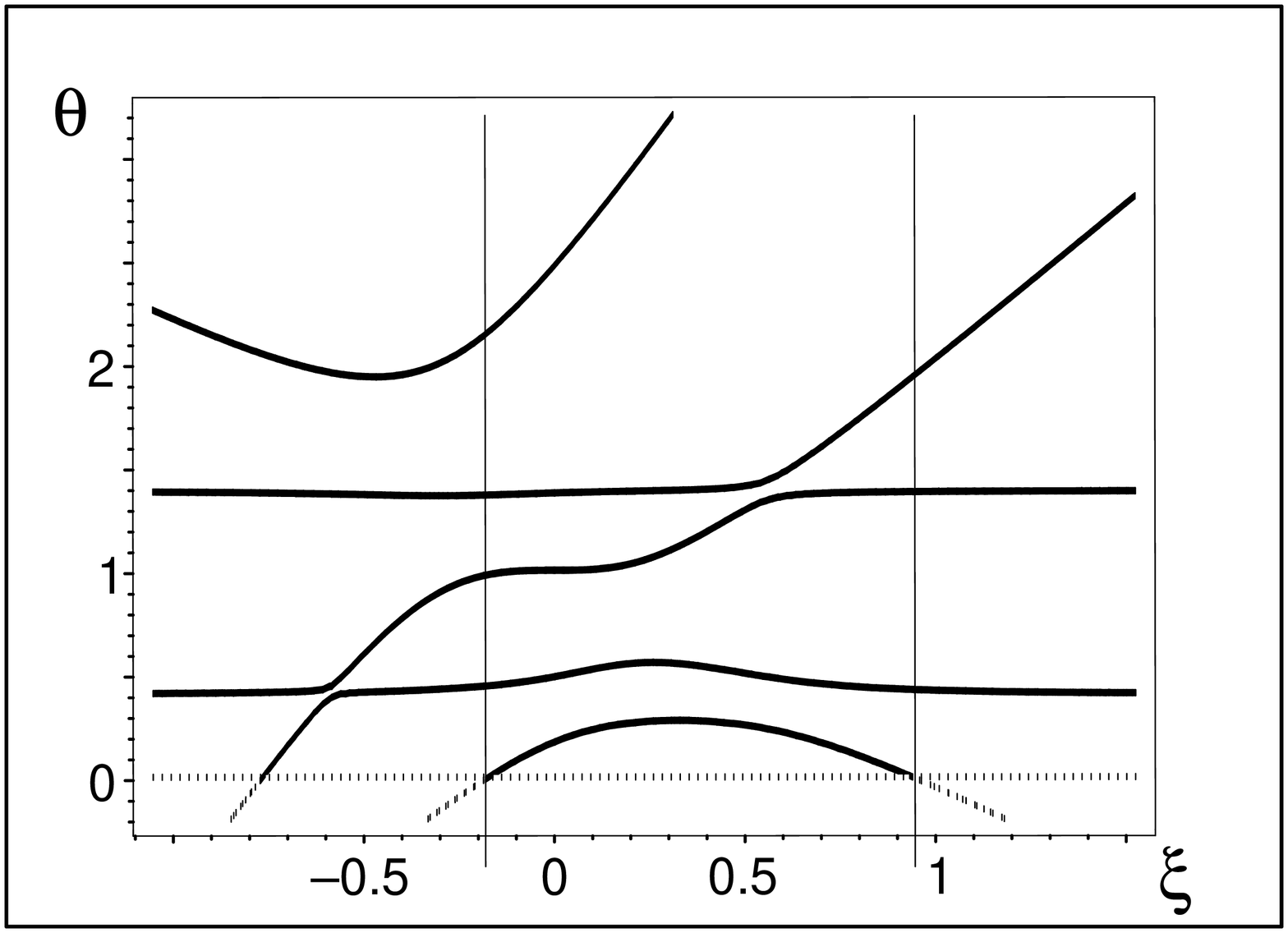,angle=270,width=0.6\textwidth}
\end{center}                         
\vspace{-2mm} \caption{Eigenvalues $\theta=\theta_j(\xi)$ of the
candidate (\ref{sv}) for the Hermitizing metric, assigned to
Hamiltonian (\ref{5dim}) at $r=1/2$ and $s=0$. The vertical lines
mark the (open) interval of $\xi$ in which $\Theta$ remains positive
definite. \label{fidmet}}
\end{figure}


%


Naturally, there emerge also several qualitative differences.
Firstly, in the two illustrative Figs.~\ref{fidm715} and
\ref{fidm7100} it is not too well visible that the absolute values
of the minimal and maximal energies grow linearly with the unlimited
decrease of  $s$. What is better visible in these pictures (where we
have $M=3$) is the survival of the coexistence of the
$s-$independent (and real) eigenvalue triplet with the not always
real quadruplet which varies with $s$.  In the non-asymptotic domain
of $|s| \ll \infty$ these functions of $s$ may possess the real EP
singularities, i.e., they may intersect their $s-$independent
partners and also - pairwise - complexify.

From the point of view of unitary quantum mechanics of stable
systems one only has to pay attention to the physical parametric
subdomains ${\cal D}_j$ in which the spectrum remains all real and
non-degenerate. At the fixed values of $r$ and $q$ these subdomains
become intervals of admissible $s$. We may find three of these
intervals in Fig.~\ref{fidm715}, and four of them forming the two
adjacent pairs separated by a non-empty gap in Fig.~\ref{fidm7100}.

\subsection{The domains of positivity of the metrics\label{hybe}}

The weakest point of the recurrent construction of the
arbitrary-dimension metrics as described in section \ref{sI} lies in
the necessity of a purely empirical verification of the positivity
of all of the eigenvalues of a given $\Theta^{(candidate)}$. The
procedure remains numerical and may be sampled using the $N=5$
Hamiltonian
 \be
 \hat{H}^{(MA)}(r,s)=\left[ \begin {array}{ccccc}
 0&-1&r&0&0\\\noalign{\medskip}-1&0&-1+s&0&0\\
 \noalign{\medskip}0&-1&0&-1+s&r\\\noalign{\medskip}0&0&-1&0&-1
 \\
 \noalign{\medskip}0&0&0&-1&0\end {array} \right]\,.
 \label{5dim}
 \ee
Naturally, a fully general discussion would be too long. Let us
therefore illustrate the whole approach and a few of its most
relevant aspects via a single sample of the determination of the
parametric domain of  positive definiteness of the metric using the
following special sparse-matrix (SSP) choice of the ansatz with
normalization $\Theta_{11}^{(SSP)}=1$, $\Theta_{12}^{(SSP)}=0$,
$\Theta_{13}^{(SSP)}=\xi$ and
$\Theta_{14}^{(SSP)}=\Theta_{15}^{(SSP)}=0$.

After one applies the above-described recurrent algorithm the
candidate for the metric $\Theta=\Theta^{(SSP)}(r,s,\xi)$ is
obtained in closed and still sufficiently compact form. In an {\it a
posteriori\,} test this matrix proves positive definite for a broad
range of its parameters (an illustrative sample is shown in
Fig.~\ref{fidmet}).

A marginal disadvantage of such a purely pragmatic and
non-systematic demonstration of the acceptability of the model in
quantum mechanics is that although the fully non-numerical form of
the resulting closed-form metric was still not too complicated, its
explicit display would not certainly fit in a single printed page.
For the presentation purposes we, therefore, choose and fixed
parameters $r=1/2$ (as usual) and $s=0$ (taken as lying safely
inside one of the above-discussed physical, real-energy domains
${\cal D}$). This already enables us to display a representative
sample of the matrix of the metric quite comfortably,
 \be
 \Theta^{(SSP)}(1/2,0,\xi)=\left[ \begin {array}{ccccc} 1&0&{\it {\xi}}&0&0\\
 \noalign{\medskip}0&1+{\it {\xi}}&-1/2&{\it {\xi}}&-1/2\,{\it {\xi}}\\
 \noalign{\medskip}{\it {\xi}}&-1/2
&1+{\it {\xi}}&-1/2-1/2\,{\it {\xi}}&1/4+{\it
{\xi}}\\\noalign{\medskip}0&{\it {\xi}}&-1/2-1/2\,{\it
{\xi}}&5/4+{\it {\xi}}&-1-1/2\,{\it {\xi}}
\\\noalign{\medskip}0&-1/2\,{\it {\xi}}&1/4+{\it {\xi}}&-1-1/2\,{\it {\xi}}&5/4
+1/4\,{\it {\xi}}\end {array} \right]
  \label{sv}
 \ee
For this particular example we arrived at the following last result.

\begin{prop}

For every Hamiltonian of Eq.~(\ref{5dim}) with two free parameters
which lie inside such a physical domain ${\cal D}$ which contains
$r=1/2$ and $s=0$ there exists a non-empty one-parametric family of
the above-specified sparse-matrix Hermitizing metrics
$\Theta^{(SSP)}(r,s,\xi)$ which are positive definite in a non-empty
interval ${\cal F}$ of eligible parameters $\xi$.

\end{prop}

\bp

For the special model in question the positivity of the eigenvalues
$\theta_j=\theta_j(r,s,\xi)$ of matrix $\Theta^{(SSP)}(r,s,\xi)$ has
an easy proof. It relies on the smoothness of the
parameter-dependence of the roots of the secular equation, say, in
the vicinity of zero $\xi=0$ and of the pre-selected values of
$r=1/2$ and $s=0$. Graphically, this feature of the eigenvalues of
our metric candidate is supported by Fig.~\ref{fidmet}. Strictly at
$\xi=0$, $r=1/2$ and $s=0$ the explicit construction of the secular
polynomial equation
 $$
 \det [\Theta^{(SSP)}(1/2,0,0)-\theta\,I]=-{\frac {9}{32}}+
 {\frac {181}{64}}\,{\it \theta}-{\frac {547}{64}}\,{{\it \theta}}^{2}+
 \frac{21}{2}\,{{\it \theta}}^{3}-\frac{11}{2}\,{{\it \theta}}^{4}+{{\it \theta}}^{5}
 =0\,
 $$
supports the exact localization of one of its roots (viz., the
``middle'', closed-form one, $\theta_3=1$). Then, the reduced
secular equation
 $$
 64\,{{\it \theta}}^{4}-288\,{{\it \theta}}^{3}+384\,{{\it \theta}}^{2}-163\,{\it
\theta}+18
 =0
 $$
is, in principle, non-numerical. In practice, it is sufficient to
evaluate the remaining four roots numerically, arriving at the
values
 $$
         0.1704659382, 0.4862291155, 1.374374593, 2.468930353
 $$
which are all safely positive.

\ep

%

\section{Discussion \label{sV}}


In many ${\cal PT}-$symmetric quantum models of bound states
the physical Hilbert-space metric $\Theta$ acts at a
distance. Its most
conventional, kinematically local
representation $\br x|\Theta|y\kt \sim
\theta(x)\delta(x-y)$
is generalized to a non-local expression \cite{alocality}.
This means that there is no physical reason for keeping the
dynamical-input Hamiltonians
local.
In our present paper we supported this
abstract observation of admissibility of non-local interactions
by a concrete constructive
support.
Using the coordinates in a discrete-lattice grid-point approximation
$x \to x_k$ with $k = 1, 2, \ldots, N = 2M+1$ we proposed a family of
Hamiltonians $H=T+V$ in which the interaction $V$ was strongly non-local
but
still user-friendly.
In particular,
the wave functions
as well as Hermitizing  metrics proved
obtainable by algebraic means. Now, we only intend to add
a few complementary comments.


Firstly, let us mention that although certain features of
non-locality (i.e., in some sense, of an ``action at a distance'')
are deeply encoded in the very formalism of quantum theory, many of
their concrete manifestations look suspicious. They evoke doubts
which may be best sampled by the famous EPR ``paradox'' \cite{EPR}
in which the (indeed, deplorable!) incompatibility of quantum
mechanics (of systems with the finite number of degrees of freedom)
with the kinematical principles of special relativity was
interpreted as an obvious ``disproof of completeness'' of the former
theory. The same old locality-involving misunderstanding reemerged,
again, in the above-mentioned Jones' papers \cite{Jones,Jonesdva} as
well as in a very recent new round of the revitalized discussion in
which the ``superluminal signaling'' argument was targeted directly
against PTSQT. According to the authors, their results ``kill any
hope of ${\cal PT}-$symmetric quantum theory as a fundamental theory
of nature'' \cite{Lee}.

It is necessary to admit that our present study was partially
provoked also
by the latter provocative conclusion.
Via the description of
our dynamically manifestly non-local models we decided to
contradict the statement by
emphasizing that
the currently accepted
formalism of quantum theory
may
really
appear strongly counterintuitive
and deeply non-local, especially
in some of its
less common implementations.
So
in support of the
acceptability of similar models
we offered an exactly solvable example.

In the light of the various intuitive perceptions of the concept of
locality (which form, in fact, a very frequent core of similar
misunderstandings), the specific PTSQT implementation of the
abstract formalism of quantum theory is particularly vulnerable,
indeed. One of the reasons is even purely historical: the variable
$x$ in the traditional benchmark PTSQT examples (\ref{tradi}) is
often called ``coordinate'' in spite of  {\em not being even real}
in general. In fact, its asymptotic complexity is a strict {\em
necessary} condition of the reality of the {\em observable}
bound-state energies for {\em all} of the toy-model potentials
$V(x)=-(ix)^\delta$ of Ref.~\cite{BB} whenever one selects
$\delta\geq 4$.

Another reason of an enhanced sensitivity of PTSQT to
misunderstandings may be found to lie in a innovative double meaning
of non-locality in this context. Indeed, in addition to the
traditional dynamical nonlocalities resulting, say, from a
replacement of the conventional potential $V(x)$ by its non-local
(i.e., e.g., momentum-dependent)  alternatives, one also has to deal
with the purely kinematical non-localities as caused by the
necessity of reconstruction of the inner product with respect to
which a given  ${\cal PT}-$symmetric Hamiltonian $H \neq H^\dagger$
would be made Hermitian. In this setting the readers are advised to
re-read, once more, Appendix A and, in particular, its application-
and physics-oriented parts in which $H$ describes a heavy atomic
nucleus (composed, as we know, of fermions) while using, highly
counterintuitively, the language of bosons.

%
%
%
%
%

In the broader context of quantum physics, naturally, the present
demonstration of the theoretical as well as practical acceptability
of a combination of the nonstandard (i.e., dynamically non-local)
interactions $V$ with the nonstandard (i.e., kinematically
non-local) representations ${\cal H}^{(S)}$ of the physical Hilbert
spaces of quantum states may be perceived as an encouragement. We
believe that our present constructive mathematical results might be
also read as opening new perspectives in the phenomenological model
building.

\newpage

\section*{Appendix A. ${\cal PT}-$symmetric quantum theory {\it in nuce}}



What is shared by the majority of textbooks on quantum mechanics is
the preference of representations  ${\cal H}^{(P)}$ of the physical
Hilbert space of states in a mathematically most friendly form, say,
of the space of square-integrable functions $L^2(\mathbb{R})$ or
$L^2(\mathbb{R}^3)$ etc. In the conventional quantum theory all of
the latter spaces could be also collectively denoted by an
alternative dedicated symbol ${\cal H}^{(F)}$ in which the
superscript stands for ``friendly''. In such a rigid setting the
bound-state Hamiltonians with real spectra must be necessarily
assumed self-adjoint of course. For our present purposes, let us
denote them, say, by a dedicated symbol $\mathfrak{h}$.

In the PTSQT context one weakens the assumptions and requires that a
given Hamiltonian (let us denote it by a different symbol $H$)
ceases to be self-adjoint in the preselected friendly Hilbert space.
Then, one has to employ a new theoretical framework called, in our
compact review \cite{SIGMA}, a three-Hilbert-space (THS)
representation of quantum theory. In this new framework the friendly
space becomes unphysical and, in terms of physics, ``false'', with
the consequence that ${\cal H}^{(F)} \neq {\cal H}^{(P)}$. The
third, ``standard'' Hilbert space ${\cal H}^{(S)}$ representing
physics must be introduced as another option which remains {\em
different} from the preceding two. Thus, the third, S-superscripted
Hilbert space ${\cal H}^{(S)}$ is, in general, {\em constructed} as
unitarily equivalent to  ${\cal H}^{(P)}$. One can summarize that in
the light and from the time of the earliest applications of the idea
by physicists in 1956 \cite{Dyson} the introduction of the third,
S-superscripted physical Hilbert space is sufficiently strongly
motivated by making the calculations as well as the interpretations
of the results perceivable simpler in technical sense (cf. also
\cite{Geyer}).

The requirements of the unitary
equivalence of ${\cal H}^{(S)}$ with ${\cal H}^{(P)}$ and, in
parallel, of the non-equivalence between the physical ${\cal H}^{(S)}$ or ${\cal
H}^{(P)}$ and the false space ${\cal H}^{(F)}$ may be easily
clarified. It is sufficient to redefine the kets $|\psi \kt^{(P)}
\in {\cal H}^{(P)}$ (treated, typically, as ``fermionic'' in
nuclear-physics applications \cite{Geyer}) as the mere images
(called, conveniently, Dyson's maps \cite{Dyson}) of certain
``bosonic'' kets which remain the same in both of the ``bosonic''
spaces, $|\psi \kt^{(F)}= |\psi \kt^{(S)}$,
 \be
 |\psi \kt^{(P)}=\Omega\,|\psi \kt^{(F)}\ \equiv \ \Omega\,|\psi
 \kt^{(S)}\,  \in {\cal H}^{(P)}
 \,,\ \ \ \
 |\psi \kt^{(F)}
   \in {\cal H}^{(F)}
 \,,\ \ \ \
 |\psi \kt^{(S)}
     \in {\cal H}^{(S)}\,.
 \ee
Naturally, whenever the Dyson's map $\Omega$ itself is admitted to
be a non-unitary operator, we may define a nontrivial Hilbert-space
metric $\Omega^\dagger \Omega = \Theta \neq I$. The required unitary
equivalence between ${\cal H}^{(P)}$ and ${\cal H}^{(S)}$ (i.e., the
postulated coincidence of the results of the respective inner
products) will then acquire, after the mere insertions, the
following next-to-elementary metric-dependent form,
 \be
 ^{(P)}\br \psi_1|\psi_2\kt^{(P)} =\
 ^{(F)}\br \psi_1|\Theta|\psi_2\kt^{(F)}\,.
 \label{netrif}
 \ee
The latter formula should now be interpreted as the definition of
the physical, correct inner product in ${\cal H}^{(S)}$.
Alternatively, we may read the latter formula as a metric-mediated
representation of the correct, S-superscripted inner product when
represented inside auxiliary space ${\cal H}^{(F)}$. Indeed, for the
purposes of the mere constructive considerations, one does not need
to work with the third, S-superscripted space at all. After an
explicit use of metric $\Theta$, all of the necessary formulae may
be pulled back to the manifestly unphysical, F-superscripted space
${\cal H}^{(F)}$. In the resulting THS setting the Hamiltonian
becomes re-interpreted as safely self-adjoint in the correct,
S-superscripted space. Equivalently, one may write $H =H^\ddagger \
\equiv\ \Theta^{-1} H^\dagger\,\Theta$ or
$H=\Omega^{-1}\mathfrak{h}\Omega$ (remember that
$\mathfrak{h}=\mathfrak{h}^\dagger$) and call operator $H$
self-adjoint with respect to the metric-mediated inner product, or
crypto-self-adjoint when studied inside the most friendly auxiliary
and unphysical F-superscripted Hilbert space.

As a consequence, the unitarity of the evolution in the physical
Hilbert space ${\cal H}^{(S)}$ becomes guaranteed by the
quasi-Hermiticity property (\ref{nelamb}) of the Hamiltonian which
is {\em manifestly non-Hermitian} in the ``false'' Hilbert space
${\cal H}^{(F)}$. The same quasi-Hermiticity feature must
necessarily characterize all of the other candidates $\Lambda$ for
acceptable physical observables \cite{Geyer},
 \be
 \Lambda^\dagger \Theta=\Theta\,\Lambda\,.
 \label{lamb}
 \ee
In multiple applications, a decisive appeal of such a THS
representation terminology (cf. \cite{SIGMA}) is seen in the
possibility of an explicit guarantee of a sufficient technical
{simplicity} of Hamiltonians $H$ in ${\cal H}^{(F)}$ \cite{Geyer}.
In our present paper the interpretation of the whole scheme follows
Refs.~\cite{alocality} and is, therefore, different because the
argument $x$ of wave functions $\psi(x) \in L^2(\mathbb{R} \equiv
{\cal H}^{(F,S)}$ {\em does not} represent, in general, an
observable point-particle position anymore. In this sense, virtually
all ${\cal PT}-$symmetric quantum models are, by construction,
non-local in the physical representation space ${\cal H}^{(S)}$
because in the conventional delta-function basis in ${\cal H}^{(F)}$
the operator of the coordinate (if any) is generically non-diagonal
(see a few concrete examples in \cite{alocality}). Thus, one can
only conclude that in the PTSQT theoretical framework the origin of
{\em any} measurable non-locality lies, more or less inseparably, in
{\em both} of its input-information sources given by the
non-Hermitian-potential {\em dynamical} input {\em and\,} by the
independent, metric-selection-related {\em kinematical} input.

\end{document}